# Efficient Hardware Implementation of Modular Multiplier over GF ($2^m$) on FPGA

*Ruby Kumari, Gaurav Purohit and Abhijit Karmakar*
*Academy of Scientific and Innovative Research (AcSIR), CSIR-CEERI Campus,*
*CSIR – Central Electronics Engineering Research Institute, Pilani, India*

## Abstract

Cryptography ensures confidentiality and provides countermeasures against cyber-attacks on sensitive information. Among cryptographic algorithms, elliptic curve cryptography (ECC) has emerged as the dominant public-key protocol, with NIST standardizing parameters for binary field GF($2^m$) ECC systems. This work presents a hardware implementation of a Hybrid Multiplication technique for modular multiplication over the binary field GF($2^m$), targeting NIST B-163, 233, 283, and 571 parameters. The design optimizes the combination of conventional multiplication (CM) and Karatsuba multiplication (KM) to enhance elliptic curve point multiplication (ECPM) in cryptographic systems. In our case the key innovation lies in using CM for smaller operands (up to the optimum value of 41 bits for 163) and KM for larger ones, after the optimal point for CM. This approach reduces computational complexity and enhances efficiency, especially when operands exceed the size of the irreducible polynomial defined by NIST standards. The design is evaluated in three key areas**:** 1. Resource Utilization**:** For m=163, the hybrid design uses 6,812 LUTs, a 39.82% reduction compared to conventional methods. For m=233, the hybrid approach reduces LUT usage by 45.53% and 70.70% compared to overlap-free and bit-parallel implementations, respectively. 2**.** Delay Performance**:** For m=163, the design achieves a delay of 13.31ns, improving by 37.60% over bit-parallel implementations and providing competitive performance compared to Montgomery method (11.70 ns), but with better resource efficiency. For m=233, it maintains consistent performance with a 13.39 ns delay. 3. Area-Delay Product **(**ADP)**:** For m=163, the design achieves an ADP of 90,860, outperforming the bit-parallel (75,337) and digit-serial implementations (43,179). For m=233, it demonstrates a 16.86% improvement over the overlap-free design and a 96.10% improvement over bit-parallel implementations. The experimental results show that the proposed hybrid multiplication technique significantly improves speed, hardware efficiency, and resource utilization, making it ideal for elliptic curve cryptography (ECC) and cryptographic hardware systems.

## I. Introduction

Cryptography ensures confidentiality, data security, and authentication in various applications like communication devices [1], autonomous vehicles [2], Internet of Things (IoT) [3] [4], and healthcare [5]. It typically involves two types of techniques: symmetric-key and public-key cryptography[6]. Public-key cryptography enables secure communication between parties without prior shared secrets [7], requiring key establishment and digital signatures. Notable examples include Diffie–Hellman [8], RSA [9], ElGamal [10], and elliptic curve cryptography (ECC) [11]. ECC is particularly favored for its strong security relative to shorter key sizes and efficient implementation [12],[13].

With the increasing reliance on information technology across various sectors, the risks associated with data security are also rising. The prevalence of cyberattacks makes it essential to secure sensitive data to prevent unauthorized access, data breaches, and identity theft. Cryptographic systems play a vital role in ensuring the safety and security of information [14]. Homomorphic authentication allows computations on encrypted data while maintaining security. A critical aspect of homomorphic encryption is large multiplication, particularly in operations like finite field multiplication. This operation is fundamental in several fields, including digital signal processing, coding theory, and cryptography. Large multiplication enables deeper levels of computation (greater multiplicative depth) while ensuring the security of data during both encryption and processing. To achieve the necessary multiplicative depth and maintain security, the operand size must be increased. However, increasing the operand size also necessitates more efficient multiplication techniques to strike a balance between security and performance. Consequently, the efficiency of the multiplier specifically, its speed versus cost becomes crucially important [9].

Many researchers worked on multiplication including classic multiplication (CM) [15], Karatsuba Multiplication (KM) [16], Overlap-free KM and Bit parallel multiplier [15], [17] [18]algorithms. Each of these n-bit multipliers differs significantly in their approach, complexity, and efficiency. The CM has O ($n^2$) complexity for a set of *n-1* degree polynomials [19]. With larger bit sizes, the quadratic growth of operations increases simultaneously making CM less efficient. Karatsuba multiplication offers a more efficient divide-and-conquer approach, reducing the number of multiplications and improving its complexity to *O* ($n^{1.58}$). Though KM reduces the number of multiplications, its recursive structure introduces a significant latency which lowers its speed and thereby performance. Contrary to KM, the Overlap-free Karatsuba algorithm (OKA) [16], [17], [20] is designed to avoid data overlap during the intermediate step improving the combinational delay introduced by the modulo-adders. This improves the complexity of KA ($3\log_2(n)-1$) Tx to ($2\log2(n) - 1$) Tx in OKA, where Tx is the processing delay of a modulo-adder.

Basically, researchers focused on many other different types of efficient bit-parallel multipliers using polynomial basis normal basis (NB) and non-conventional basis [14], [21]. These multipliers are efficient ones reducing delay up to (1 + log2(n − 1)) $T_X$ by combining irreducible all-one polynomial with three-term KA. They exploit parallelism at the bit level, speeding up the multiplication process by performing parallel operations simultaneously at the cost of a higher HW area. Therefore, there is an ample need for an area and power-efficient multiplier for large number multiplications yet simple at the same time. This paper proposes a new approach integrating two most promising multipliers i.e., CM with KM with modular reduction by the irreducible polynomial resulting in an efficient hybrid KM modular multiplier. The hybrid method finds an optimal operand size among chosen NIST Binary Field Curves (B-163, B-193, B-233, B-283, B-571) and improves the complexity to 3 + log2(n/2) performing superior to all the existing quadratic and sub-quadratic multipliers tailored for the $GF(2^m)$ multiplications. This approach is suitable for critical applications viz. SCA-resilient algorithms as well as cryptographic applications where efficient hardware acceleration is required [23].

The rest of the paper is organized as follows, Section II reviews the finite field multipliers, lager number multiplication, reduction polynomial, and related work. Section III presents the details of the FPGA implementation procedure and results for the algorithms mentioned above and further introduces the proposed hybrid multiplication strategy. Performance and device utilization as well as comparison with similar and relevant works are discussed in Section IV. Section IV, finally concludes this article.

## II. Background

### A. Finite Field Multipliers

A finite field $GF(2^m)$ consists of binary polynomials with coefficients from {0,1}. Each element of $GF(2^m)$ is represented by a polynomial of degree lesser than m, as (1)

$$a(x) = a_{m-1} * x^{m-1} + a_{m-2} * x^{m-2} + \cdots + a_1 * x + a_0 \tag{1}$$

$$b(x) = b_{m-1} * x^{m-1} + b_{m-2} * x^{m-2} + \cdots + b_1 * x + b_0 \tag{2}$$

where the coefficients $a_i$ & $b_i \in GF(2)$. Polynomial multiplication involves coefficientwise multiplication of both first and second polynomial to result in $c(x)$ as (3). This leads to an operation with time complexity of O ($n^2$) for degree m polynomials.

$$c(x) = a(x) * b(x) \tag{3}$$

To ensure the result (3) in $GF(2)$, i.e. the resulting polynomial's degree, lesser than $m-1$ and it is must to perform a modular reduction by the irreducible polynomial $p(x)$. Usually, the chosen p(x) has property that it cannot be factored into polynomials of lower degree with coefficients in the same field. So, all operations including addition and multiplication are reduced with an p(x) of degree m to ensure the result is an element of GF($2^m$) as (4)

$$c'(x) \ = C(x) \ mod \ P(x) \tag{4}$$

$$\begin{aligned} c(x) \quad &= c_{2m-2}x^{2m-2} + \cdots + c_m x^m + c_{m-1}x^{m-1} + \cdots + c_1 x + c_0 \\ &\equiv (c_{2m-2}x^{m-2} + \cdots + c_m)r(x) + c_{m-1}x^{m-1} + \cdots + c_1 x + c_0 \quad \big(\text{mod } P(x)\big). \end{aligned}$$

The process of coefficientwise multiplication is relatively simple than other multiplication technique for binary fields but possess overheads due to bit-level manipulations involved in it. To get in depth lets us see the basic 2-bit and 4-bit building blocks of multipliers to extend the same for n-bit multipliers later in the next section.

**B. Large Numbers Multiplications**

The conventional algorithm for binary CM uses 2-bit or 4-bit multiplier block using AND, and XOR logics as shown in figure 1. Figure 1 shows a typical hardware implementation of a 2-bit CM. The circuit have 2 input bits $a_0$, $b_0$, & $a_1$, $b_1$ and three outputs bits $c_0$, $c_1$ & $c_2$. The combinational architecture uses AND logics for multiplication and XOR logics for bitwise operations. The output $c_0$ and $c_2$ are extracted by and logics whereas $c_1$ is bitwise addition of multilevel partial products of inputs. Here the example of simple 2-bit multiplier is implemented using 4 AND gates and only 1 XOR gate.

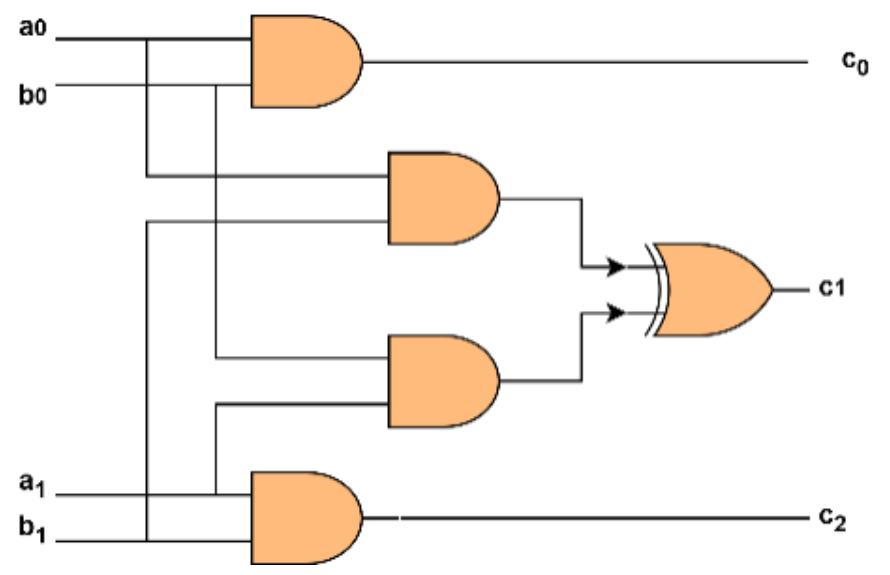

Fig. 1. (a) Hardware implementation of 2-bit binary polynomial

The similar logics can be extended for 4-bits as well as for m-bits CM. For m-bit CM, equation (2, 3) is modified and the gate count will get increased as (5, 6),

$$CM_{XOR}(m) = (m-1)^2 \tag{5}$$

$$CM_{AND}(m) = (m)^2 \tag{6}$$

Where $CM_{XOR}(m)$ and $CM_{AND}(m)$ are the total number of XOR and AND logics respectively. As the logic count increase the overheads due to bit-level manipulations get increases. Assuming ideal hardware condition and good signal strength i.e. no intermediate register or memory are used, the delay of the 2-bit CM will be as (7)

$$T_{CA}(2) = T_a + T_x \tag{7}$$

where $T_x$ and $T_a$ are the delay of an XOR and an AND gate respectively. Similarly, for m- bit multiply, (m − 1) th term observes maximum path delay and Tx observe logarithmic trend as shown in (8) [ref]

$$T_{CA}(m) = T_a + log_2(m)T_x \tag{8}$$

The multiplier efficiency depends upon its total delay and HW area. However, these multipliers are very ineffective for larger operand sizes because the area does not scale as the input operand size grows as shown in Figure m-bit. This challenges designers to develop new architecture to keep balance in both areas and delay efficiency keeping design simple. One widely known algorithm KM offers significant performance improvements over traditional CM for large operands. To describe the KM algorithm, input $a_i$ & $b_i$ as (1) will get split as (x) to produce $c(x)$, as the product of degree ≤ 2m-2 as in (11) [22]

$$a(x) = x^{\frac{m}{2}}\left(x^{\frac{m}{2}-1}a_{m-1} + \cdots + a_{\frac{m}{2}}\right) + \left(x^{\frac{m}{2}-1}a_{\frac{m}{2}-1} + \cdots + a_0\right) = x^{\frac{m}{2}}A_H + A_L \tag{9}$$

$$b(x) = x^{\frac{m}{2}}\left(x^{\frac{m}{2}-1}b_{m-1} + \cdots + b_{\frac{m}{2}}\right) + \left(x^{\frac{m}{2}-1}b_{\frac{m}{2}-1} + \cdots + b_0\right) = x^{\frac{m}{2}}B_H + B_L \tag{10}$$

$$C(x) \ = \ x^m A_H B_H + (A_H B_L + A_L B_H)x^{\frac{m}{2}} + A_L B_L \tag{11}$$

The algorithm requires three multiplications and four additions per recursive step at each level. It can be observed with the 2-bit Karatsuba multiplication circuit design. It uses four XOR logic gates as adders and three AND gates for multiplication shown in Figure 2(a). In this circuit, the input operands are denoted as $a_0, a_1$ and $b_0, b_1$, and the final product of the circuit are $c_0, c_1,$ *and* $c_2$, bits. The 4-bit KA circuit uses three 2-bit KA and six, 2-input XOR gate with the overlap logic as highlighted (ii) in Figure 2 (b).

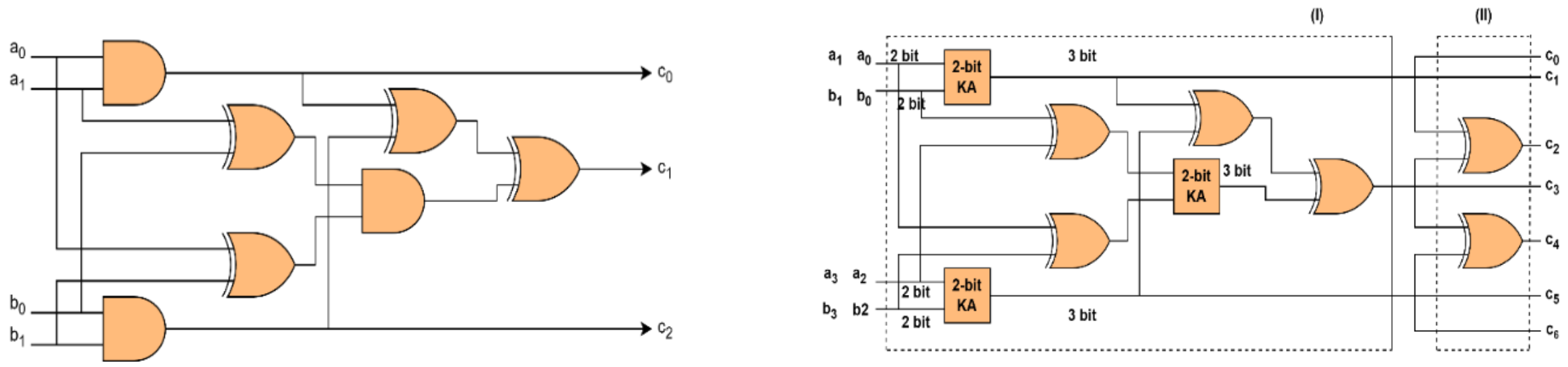

Fig. 2. (a)  Schematic realization of 2-bit Karatsuba multiplication (b) Schematic realization of 4-bit Karatsuba multiplication

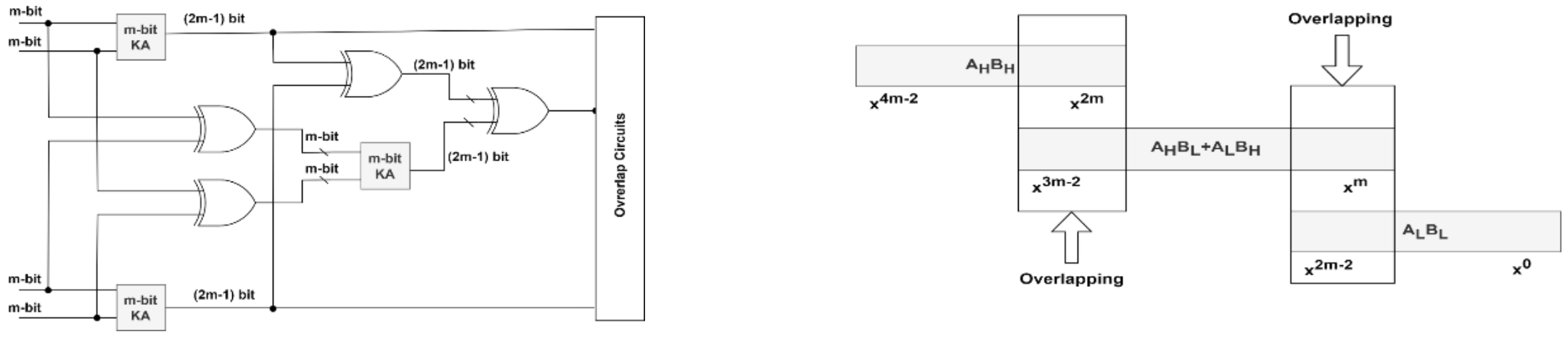

Fig. 3. (a) Schematic realization of m-bit Karatsuba multiplication algorithm (b) Schematic realization of Karatsuba multiplication overlap circuit [23]

The figure 3(a) extends KA to an m-bit Karatsuba multiplier for multiplying m bits larger number. It shows two m-bit inputs on the left, each split into smaller portions for multiple sub-multiplications of (m/2)-bit units. The XOR gates combine partial products through addition and subtraction. The interconnected logic gates implement the divide-and-conquer approach of the Karatsuba algorithm, reducing the number of single-digit multiplications by breaking the problem into three smaller multiplications of size m/2, along with a few addition operations. The final (2m-1) bit product is produced as intermediate output. These intermediate results are then combined by an overlap circuit. The overlap circuit has an XOR logic structure that aligns and merges the bits from the intermediate outputs to produce the final (2m-1)-bit product as in Figure 3(b).

The KM reduces hardware area but impacts speed performance due to the delay introduced by the modulo adders in its architecture. The Overlap-free Karatsuba algorithm (OKA) algorithm aims to improve the combinational path delay introduced by these modulo adders in the critical path while multiplication [24]. The OKA results in a considerable reduction in the theoretical delay in comparison to KA. Figure 4 shows the 4-bit OKA circuit design using 2-bit KA as shown earlier. Here four XOR logic gates as adders and three AND gates for multiplication are shown in Figure 2. In this circuit, the input operands are denoted as $a_0, a_1$ and $b_0, b_1$, and the final product of the circuit are $c_0, c_1,$ *and* $c_2$, bits.

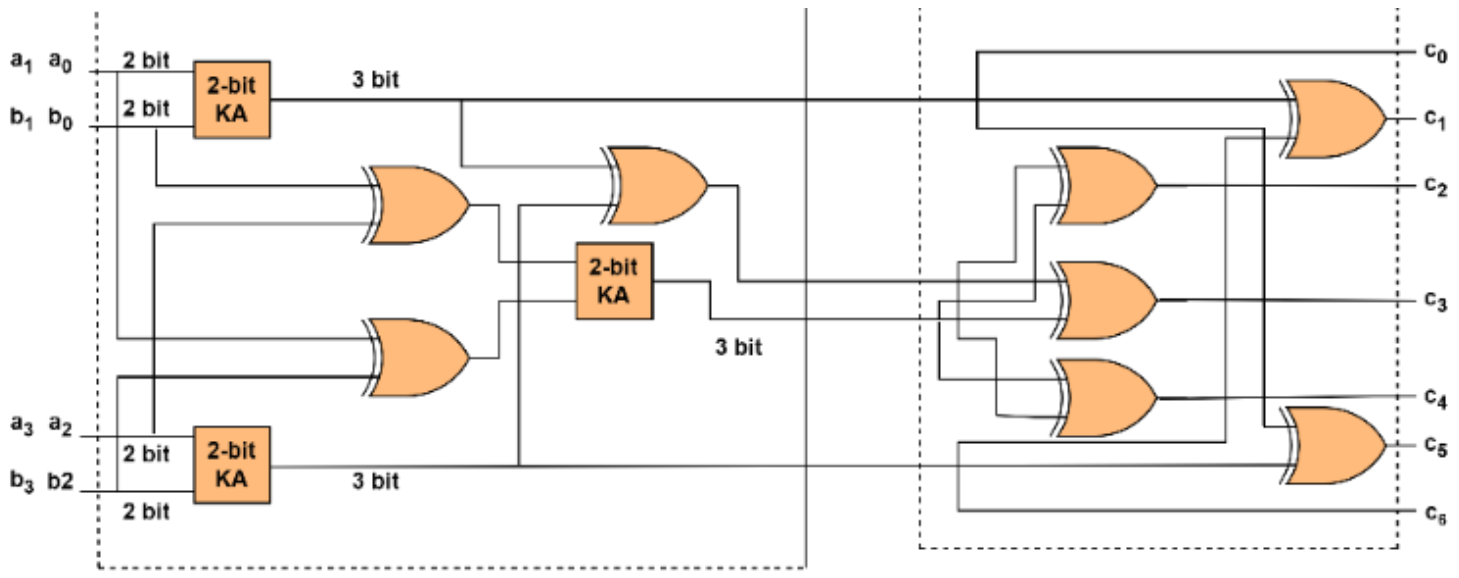

Fig. 4. Schematic realization of 4-bit Overlap free Karatsuba multiplication

Table 1 Complexity analysis of different types of multiplication algorithms

| Scheme | AND | XOR | Total |
|---|---|---|---|
| CM | $m^2$ | $(m-1)^2$ | $T_a + log_2^m\, T_x$ |
| KM | $m^{log_2 3}$ | $6m^{log_2 3} - 8m + 2$ | $T_a + 3[log_2(m-1)]T_X$ |
| OKA/OBS [24] | $m^{log_2 3}$ | $6m^{log_2 3} - 8m + 2$ | $T_a + 2[log_2(m-1)]T_X$ |
| Bit-parallel[25] | $\frac{3m^2 + 2m - 1}{4}$ | $\frac{3m^2 + 24m + 8n + \delta}{4}$ | $T_a + 3[log_2(m+1)]T_X$ |
| Digit Serial[26] | $3N_A(\frac{m}{2})$ | $3N_X\left(\frac{m}{2}\right) + 7\left(\frac{m}{2}\right) - 3$ | $T_A(m) = T_A\left(\frac{m}{2}\right), T_X(m) = T_X\left(\frac{m}{2}\right) + 3$ |
| Hybrid [23] | $3^k(\frac{m}{2^k})$ | $3^k\left(\frac{m}{2^k} - 1\right)^2 + 8n\left(\left(\frac{3}{2}\right)^k - 1\right) - 2(3^k - 1)$ | $T_a + 3kT_X + log_2\left(\frac{m}{2^k}\right)\ T_X$ |

*k=1 for hybrid design

### C. Related work

Using these multipliers researcher have explored various architecture and their hardware implementations for binary fields. KM becomes obvious choice as one can take leverage of its recursive nature to achieve parallelism and reduce critical path delays. Samanta et al [27]. proposed a modified KM (MKM) for 8-bit operands, optimizing product term splitting to reduce operation delay. The MKM architecture is simple, offering a good trade-off between area and speed, making it suitable for hardware applications for circular convolution and crypto algorithms.

The field multiplication, which is crucial to field arithmetic, consists of a regular binary polynomial multiplication followed by a reduction modulo f(x). Usually, bit-parallel multipliers can be implemented using a product matrix that combine the above two steps together. Yin Li et al [28] proposed a Chinese remainder theorem (CRT)-based hybrid bit-parallel multiplier type-I irreducible pentanomials with reduced space complexity Besides KA, the Winograd short convolution algorithm and Chinese Reminder Theorem (CRT) are other well-known divide-and-conquer algorithms, widely applied to develop sub-quadratic space complexity multipliers [28], [9]. KA-based hybrid multipliers usually require at least one more TX compared with the fastest quadratic multipliers [4], [15], where TX is the delay of one 2- 2-input XOR gate. Imaña [51] has implemented a bit-parallel polynomial multiplier with a novel splitting approach for type I irreducible polynomials. It claims the lowest delay with a balanced ATP in comparison to a similar bit parallel technique on Xilinx Artix-7 FPGA. S Arish et al [29]. discussed a bit-parallel multiplier based on the Karatsuba-Urdha algorithm. The method is tailored for combining KM with the "Urdhva-Tiryagbhyam Sutra" for unsigned binary mantissa multiplication. The design later tested on Spartan-3E and Virtex-4 FPGAs for generic applications. Digit serial: Trong-yen et al [26]. (2013) presented low complexity hybrid multiplier by combining digit-serial architecture with KM for GF(2m) multiplication. Traditionally the delay time for digit-serial multiplier is O(m/d), where m is input operand size and d is digit size which they further reduced to O(ceil(m/2d) +1).

Moslem H. et al [24]. (2021) proposed speed-optimized version of KM i.e. OKA with a hardware implementation of a finite field multiplier that achieved lower combinational delay and area-delay product compared to state-of-the-art designs. However, they have not demonstrated modular OKA design, restricting its usages towards GF(2m) multipliers limiting its flexibility [14]. KM-Hybrid [30]: Here simple Karatsuba has a small modification i.e. one bit padding in KM. The new approach claims lesser space as tested on the Virtex-4 FPGA device. Renita J. et al. (2022) presented another hybrid KM architecture as proposed as in [31] optimizing it for scalar multiplications. They realized hybrid KM-Vedic multiplier combining KM at first stage and Vedic multiplier at its second stage and tested it on AMD-Xilinx Virtex-4 and Virtex-7. Zhengzheng Ge et al. (2011) [32]introduced a hybrid multiplication technique to truncate the KM algorithm earlier. They then applied various alternative techniques, including the classic algorithm and Mastrovito multipliers, to perform the remaining n-bit computations. Their technique needs a zero padding when the degree becomes an odd after splitting. The truncation is adopted on and when m <=11(if m becomes an odd) or m<= 4 (if m becomes an even) to stop the Km iteration further [21].

Further Haining Fan et al. (12) rediscovered Montgomery's N-residue method resulting in n-term Karatsuba-like formulas as the conventional Chinese remainder theorem (CRT). This is an alternate approach who provides valuable insights into Karatsuba-like formula structures maintaining the same multiplication complexity. Zhou et al.[33] introduced another hybrid approach mixing bit-parallel with KM (KOMs) like [OKM] and described as weighted trees. Their design used common expression sharing by analysing odd-term polynomials achieving lower area-delay product than any recent bit-parallel multipliers designs. KOM design has lesser complexity makes the work relevant for ECC with prime bit-depths as recommended by the National Institute of Standards and Technology (NIST). The design later verified on both Xilinx Virtex-4 & Virtex-5 devices. Xie et al. [34] presented another Hybrid design KM for digit-serial systolic multiplication. The approach involves redundant register elimination, minimizing register sharing and later two-stage pipelining to reduce overall register complexity of the proposed design. FPGA synthesis results demonstrate that the proposed multipliers outperform existing designs in terms of area, time, and power efficiency for NIST trinomial GF ($2^m$) with p = 4. The design uses Intel Altera Stratix II device for validation. Rashid [35] presented low-area and scalable digit-digit hardware structure of the polynomial basis multiplication over a finite field F2m. The multiplier offers adjustments in clock cycles and input words on the chosen digit size making it suitable for various cryptographic applications. Further the architecture has flexibility as well as scalability allowing ease of testing for binary finite fields F2163 and F2233 on Virtex-4 and Virtex-5 FPGAs.

Eduardo et al. [36] enhanced the traditional KM algorithm for binary field multiplication in a polynomial basis. Their design integrated the modular reduction using parallel linear feedback registers (LFSR) followed by polynomial multiplication. The HW architecture is described in VHDL and synthesized on Virtex-6 FPGA device. Imaña [37] introduced a bit-parallel polynomial multiplier using a novel splitting approach for type I irreducible polynomials. In comparison, our proposed method is nearly twice as fast and has an area-delay product (ADP) that is one-third smaller. FPGA implementations on Xilinx Artix-7 confirm its lower delay and balanced area-time complexity, outperforming similar multipliers. Christoph Nagl et al. (2014) demonstrated that the Karatsuba multiplier, traditionally used for high-throughput implementations, can also be a viable option for area-constrained designs, offering a balance between performance and resource utilization [38]. However, ECC of core is constituted the multiplier. Multiplier design direct effect to the safety and performance of ECC. In the recently, many researchers to study finite filed multiplier as reference papers [3-6] proposed many different of GF(2m) multiplier which including bit-serial, bit-parallel and digit-serial structure, bit-parallel multiplier usual adopted Least Significant Bit (LSB) or Most Significant Bit (MSB) of way. Recent research has focused on optimizing hardware implementations of finite field multiplication for cryptographic applications, particularly elliptic curve cryptography (ECC).

### D. Reducible polynomial for modular reduction (Unified modular reduction method)

In general, there are two steps involved in computing the finite field modular multiplication (MM) over GF (2m). In first step the multiplication of two binary polynomials of degree ≤ m −1 obtaining c(x) of degree at most 2m −2 as shown in section XX. In second step the modular reduction is used to achieve the same degree ≤ m−1 like input operands. Performance of hardware ECC is considerably affected by the speed of modular multiplication (MM) to process finite field operations. Among the above-mentioned Multipliers KM offers significant performance improvements for large operands whereas the reduction is accelerated by precomputing the polynomials. Generally, these polynomials are irreducible and standardized by NIST organization [22]. For a given size m, the irreducible polynomial can be $P(x) = x^m + r(x)$, where $r(x)$ is a binary polynomial of degree at most $m - 1$. The result $c(x)$ modulo $P(x)$ as in (3) is obtained by taking one bit at a time, starting from leftmost bit. The reduction is accelerated by precomputing the polynomials $x^k r(x), 0 \leq k \leq W - 1$. Here polynomial can be either trinomial or pentanomial based on the curves recommended by NIST [4] as depicted in table 1, for various cryptographic applications.

Table 2 polynomials recommended by NIST in the FIPS 186-2 standard [book]

| m | Irreducible Polynomial [39] | Type |
|---|---|---|
| 6 | $x^6 + x + 1$ | Trinomial |
| 11 | $x^{11} + x^2 + 1$ | Trinomial |
| 21 | $x^{21} + x^2 + 1$ | Trinomial |
| 41 | $x^{41} + x^3 + 1$ | Trinomial |
| 82 | $x^{82} + x^8 + x^3 + x + 1$ | Pentanomials |
| 163 | $x^{163} + x^7 + x^6 + x^3 + 1$ | Pentanomials |
| 233 | $x^{233} + x^{70} + 1$ | Trinomial |
| 283 | $x^{283} + x^{12} + x^7 + x^5 + 1$ | Pentanomials |
| 571 | $x^{571} + x^{10} + x^5 + x^2 + 1$ | Pentanomials |

The reduction process is usually faster for lower operand sizes or chosen $P(x)$ itself has low-degree polynomials. The process of reduction is generalized for a chosen irreducible $P(x)$ for reducing (2) to obtain (3). For this (3) can be rewritten such that the higher order bits represent c(x) and lower m bits represent $P(x)$ as in (4). To ease the modularization process is further divided into four sub-vectors i.e., W, X, Y, and Z as in (12). Let me help you rewrite this reduction scheme for irreducible polynomials over GF($2^m$). The reduction scheme can be described as follows: For a polynomial reduction over GF($2^m$), let C' be represented as a *2m-1-bit* value that needs to be reduced. The reduction process can be broken down into four primary components (W, X, Y, and Z) that are combined to form the final reduced polynomial C.

The components are defined as:

*W*: Represents the least significant m bits of C', denoted as *C'([0, m-1])*

*X*: Comprises the most significant m bits of C', expressed as *C'([m,2m-1])*

*Y*: Represents the subset of bits from position m to 2m-1-n, shifted by $x^n$, written as *C'([m,2m-1-n])* $x^n$

*Z*: Consists of the sum of bits from position 2m-n to 2m-1, also shifted by $x^n$, expressed as *(C'([2m-n,2m-1]) + C'([2m-n,2m-1]))* $x^n$. The final reduced polynomial C' is computed as C' = W ⊕ X ⊕ Y ⊕ Z, where ⊕ represents the XOR operation in GF($2^m$). This reduction scheme effectively maps a *2m-1 bit* value to an *m-bit* result while maintaining the field properties of GF($2^m$). The process utilizes the irreducible polynomial properties to ensure the result remains within the finite field.

$$C' = C'_{[0,m-1]} + C'_{[m,2m-1]} + C'_{[m,2m-1-n]}x^n + \left(C'_{[2m-n,2m-1]} + C'_{[2m-n,2m-1]}x^n\right) \tag{12}$$

$$W = C'_{[0,m-1]}$$

$$X = C'_{[m,2m-1]}$$

$$Y = C'_{[m,2m-1-n]}x^n$$
$$Z = C'_{[2m-n,2m-1]} + C'_{[2m-n,2m-1]}x^n$$

Thus, the reduction step can be computed by the addition of four terms, with a trivial process as shown in fig 5. The entire modular multiplier with KM can be rewritten as in algorithm 1. This is unified for all the polynomials as in table 2.

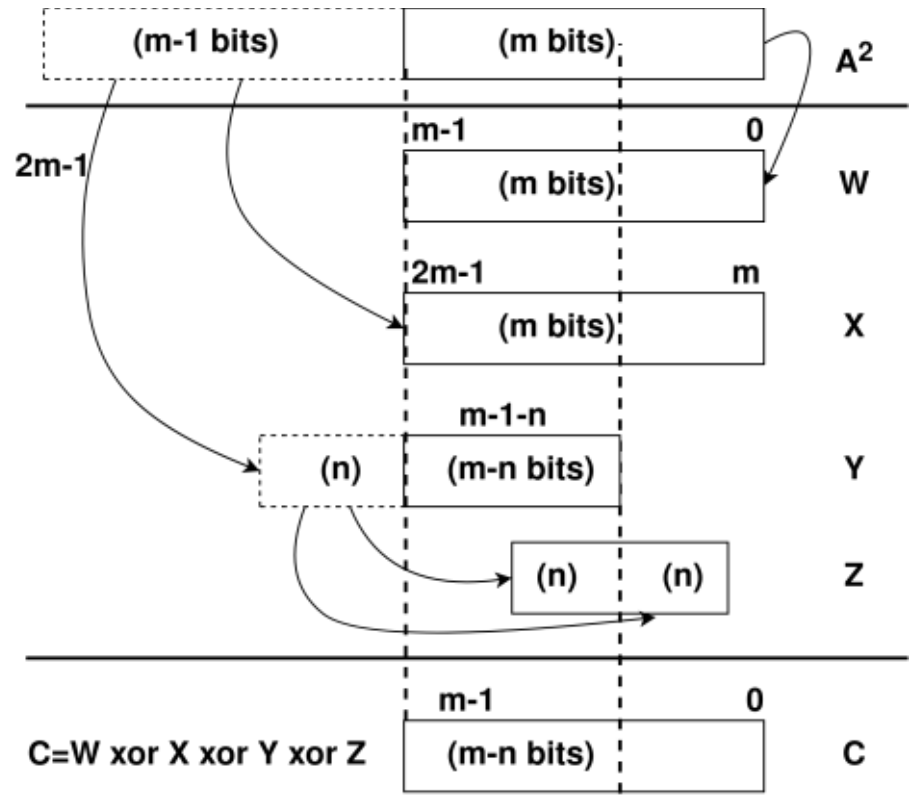

Fig. 5. Reduction Scheme [40].

The multiplication process can be computationally expensive, particularly for large numbers, as the hardware complexity increases with the size of the operands. To address this, the reduction operation ensures that the degree of the resulting polynomial is always strictly less than that of the modulus polynomial. By limiting the degree in this way, we prevent the polynomial results from becoming excessively large. We evaluated the modular multiplication using two approaches modular CM (Conventional Method) and modular KM (Karatsuba Method), to establish an efficient computation for obtaining the product C′(x) in the context of an Efficient Computing Platform Model (ECPM) application. In this case, P(x) must be fixed for a chosen m, where its highest degree is m, the same as the degree of the irreducible polynomial.

| Algorithm 1: Modular KM algorithm |
|---|
| INPUT: $A(x), B(x), P(x) \in GF(2^m)$, m=operand size<br>OUTPUT: $C'(x) = C(x) \bmod P(x)$<br>Step 1: Split $A(x)$ and $B(x)$ into lower and higher halves<br>$A(x) = x^{m/2}A_H + A_L$ & $B(x) = x^{m/2}B_H + B_L$ // L=lower, H=higher<br>Step 2: Compute $M_0, M_1, M_2$ i.e. partial products and cross term<br>$M_0 = A_L(x)B_L(x)$ ; $M_1 = A_H(x)B_H(x)$<br>$M_2 = [A_L(x) + A_H(x)] * [B_L(x) + B_H(x)] \oplus M_0 \oplus M_1$<br>Step 3: Combine<br>$C(x) = M_1x^{2m} + M_2x^m + M_0$<br>Step 4: Reduction<br>$C'(x) = [M_0(1 + x^{\frac{m}{2}}) \oplus M_2(x^{\frac{m}{2}} + x^m) \oplus M_1x^{\frac{m}{2}}] \bmod P(x)$ |

For modular CM, we applied basic polynomial multiplication followed by reduction. The resources required for hardware implementation were measured across different values of m, as shown in Table 3.a. Similarly, for modular KM, we used Algorithm 1 and varied the value of m to estimate the FPGA area requirements, which are displayed in Table 3.b. Both modular CM and KM schemes were implemented in VHDL, and the resource estimation for each scheme was carried out on a reconfigurable platform, specifically the Virtex 7 FPGA, using Vivado 2023.1 for synthesis and implementation.

Table 3a: Resource Estimation for Modular CM

| Operand Size | LUT | Delay (ns) | ADP |
|---|---|---|---|
| 4 | 7 | 4.530 | 31.71 |
| 6 | 16 | 5.008 | 85.77 |
| 8 | 32 | 5.099 | 163.17 |
| 11 | 49 | 6.363 | 311.79 |
| 21 | 185 | 8.116 | 1501.46 |
| 41 | 694 | 9.655 | 6700.57 |
| 82 | 2599 | 12.031 | 31268.57 |
| 163 | 9982 | 18.129 | 180963.68 |

Table 3b: Resource Estimation for Modular KM

| Operand Size | LUT | Delay (ns) | ADP |
|---|---|---|---|
| 4 | 7 | 5.802 | 40.61 |
| 6 | 16 | 6.002 | 96.03 |
| 8 | 146 | 7.085 | 1034.41 |
| 11 | 158 | 7.902 | 1119.43 |
| 21 | 206 | 9.083 | 1871.08 |
| 41 | 695 | 10.562 | 7340.59 |
| 82 | 2306 | 13.280 | 30623.68 |
| 163 | 7762 | 20.282 | 157428.88 |

It can be depicted easily from both the tables that the CM performs better than KM when m≤41 is lesser otherwise KM performs better. These tables interestingly reveal that designers need to carefully choose between CM and KM for a given m. Conventionally literature shows that KM usually performs better above m>96 [49]. Our primary goal is to identify the operand length where Modular CM performs optimally than KM. In fact, this can be identified just by observing Tables 3a and 3b closely. For a chosen NIST B-163, m≤41, the CM performs either better or equal to KM. On an average m<41, CM performs approximately 10-78% better in the HW area than KM. On the contrary, the KM performs better after m≥41 as marked in the green color arrow. Here color green distinguishes the scheme to be chosen over the red-marked scheme. Though at m=41 the KM has a 1 LUT lesser than CM the delay consumption is higher by 1 nanosecond. This makes the designer think of an optimal point w.r.t. area and delay tradeoff. Further one can switch to the nearest m to achieve either the best area or the best ADP to mitigate the tradeoff between area and delay. It also motivates a designer to find an appropriate m, to integrate CM and KM developing an optimal hybrid MM algorithm.

## III. Proposed Method for hybrid multiplication Strategy

The key contribution of the paper is to define an optimum split point to integrate CM at the first stage and later KM to design a new hybrid multiplier approach. CM multipliers are optimized for both area and performance at lower m. The KM has proven low latency for large number multiplication using a recursive splitting strategy that systematically breaks down the operands into smaller, more manageable components. As stated in section II, we find an optimum point i.e., size 'm' where CM is better than KM. For that one needs to take the highest power of operand and assign it as level 0 or parent node to construct a tree-like structure. Then recursively split the m to draw the leaf nodes. These leaf nodes are now at level 1 with acquiring values from set {m/2, m/2±1) and the process will go on reaching a value of m up to 2 or 3. For our ECPM case m is 163 bits and Figure 6. a, shows the tree structure with the parent node as level 0 with its highest power i.e., 163. Further, its leaf nodes of level 1 attain values i.e., m= {82,81}. These values can be assigned irrespective of their position i.e., either of them can be assigned as left or right. Now the tree grows in both directions with further recursive splitting, making two leaf nodes at level 2 with m=81 splitting into 41 & 40 and m=82 splitting into 41 & 41. Likewise, the tree terminates reaching either m= 2 or 3 value.

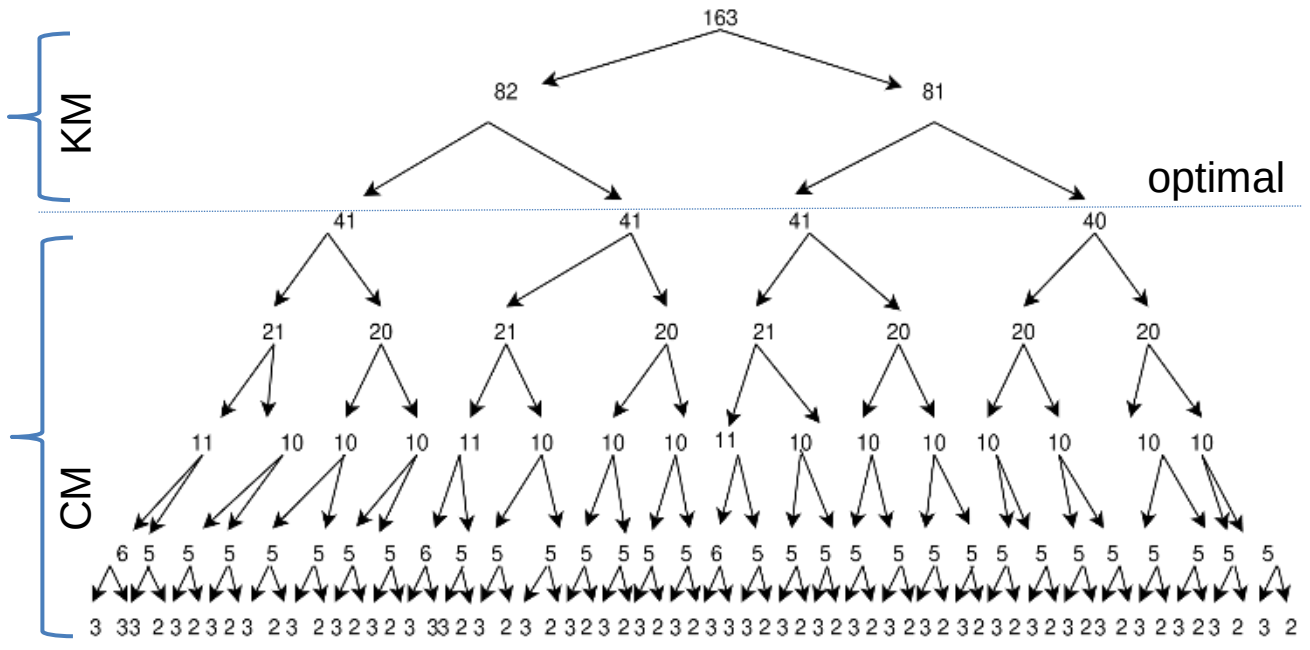

Fig.6.a Approach to deduce optimal multiplier m=163.

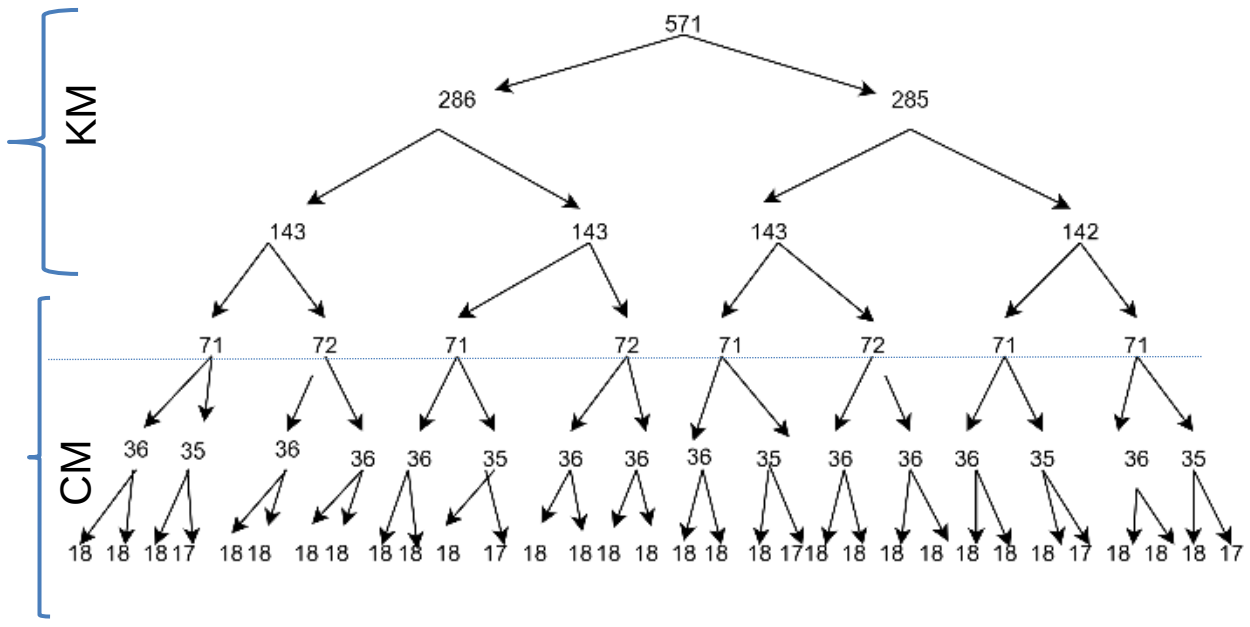

Fig.6.b Approach to deduce optimal multiplier m=571.

This hierarchical decomposition structure is fundamental to the Karatsuba algorithm's efficiency, and the similar process for NIST B-571 as in figure 6 (b) can be written empirically as:

1. Level 0: Parent (571) → Leaf1(286) + Leaf2(285)
2. Level 1: Parent (286) → Leaf1 (143) + Leaf2 (143); Parent (285) → Leaf1 (142) + Leaf2 (143)
3. Level 2: Parent (143) → Leaf1 (71) + Leaf2 (72); Parent (142) → Leaf1 (71) + Leaf2 (71)
4. Level 3: Terminal segments of CM Each 71-bit segment → 35 + 36 bits; 72-bit segment → 36 + 36 bits

The splitting process results in an optimized structure ensuring efficient resource utilization and balanced computation. Now this split gives specific m for which RTL design blocks will be written to estimate resources in terms of area, power, and delay. This will provide us optimum m up to which CM performs superior than KM. In stage 2 our obvious choice for larger m becomes KM. This strategy to estimate the resources first and find optimal m to use CM is new and explored for all the possible curves to generalize our approach. The table 4 shows the optimal points for NIST standard curves.

Table 4: Resource Estimation for Modular CM and KM

| NIST Standard | Irreducible Polynomial | Optimal Point (m) Stage I CM | Stage II KM | Application |
|---|---|---|---|---|
| 163 [41] | $x^{163} + x^7 + x^6 + x^3 + 1$ | 41 → 21 → 11 → 6 → 3 → 2 | 163 → 82 | ECPM for ECC |
| 233 [42] | $x^{233} + x^{70} + 1$ | 59 → 30 → 15 → 8 → 4 → 2 | 233 → 117 | |
| 283 [43] | $x^{283} + x^{12} + x^7 + x^5 + 1$ | 71 → 36 → 18 → 9 → 5 → 3 | 283 → 142 | |
| 571 [44] | $x^{571} + x^{10} + x^5 + x^2 + 1$ | 71 → 35 → 17 → 8 → 4 → 2 | 571 → 286→142 | |

In this paper we have proposed a hybrid Karatsuba multiplier by utilizing the conventional PM in place of an intermediary KM stage, thus requiring no further Karatsuba decompositions. The hardware implementation of KM across various operand lengths demonstrates a systematic pattern of recursive decomposition, with each variant following a carefully optimized splitting sequence. Four specific cases have been analyzed: 163-bit, 233-bit, 283-bit, and 571-bit multipliers. Then further making a hybrid design for the specified NIST Standard curve given in Table 4. The first design is shown in Figure. 8 where a 41-bit PM is utilized to achieve this. The analysis behind the obvious choice of 41 bits for our approach is to present the subsequent design depicted in Figure 8. For the 163-bit multiplier, the recursive splitting sequence follows: 163 → 82 → 41 → 21 → 11 → 6 → 3 → 2. The 233-bit multiplier demonstrates the sequence: 233 → 117 → 59 → 30 → 15 → 8 → 4 → 2. For the 283-bit implementation, the progression follows: 283 → 142 → 71 → 36 → 18 → 9 → 5 → 3. The 571-bit multiplier exhibits the sequence: 571 → 286 → 142 → 71 → 35 → 17 → 8 → 4 → 2.

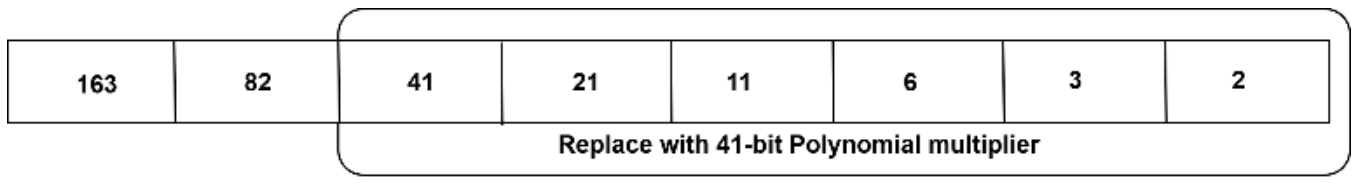

Fig. 7. Proposed structure of 163-bit hybrid Karatsuba multiplier.

Figure 7 is a block representation for the proposed hybrid method for case B-163 i.e., ECPM. The hybrid multiplier employs a unique optimization strategy where the recursive Karatsuba decomposition proceeds with 82-bits onwards and CM is considered up to 41-bits as stage I. This optimal strategy to stop at level 2 has appropriate established facts from the resource estimation tables. Similarly, for B-233 stage I i.e., CM is employed up to 59-bit and thereafter KM for subsequent operations as shown in Figure 8.

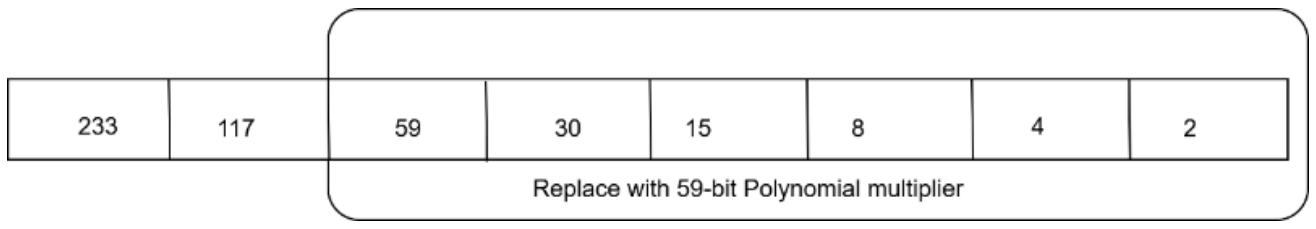

Fig. 8. Proposed structure of 233-bit hybrid Karatsuba multiplier.

The block diagram of the structure as shown in Fig. 8 can be realized using digital combinational logic. The binary hybrid KM architecture requires modulo adders i.e., XOR gates & AND gates as primary components for the realization of GF(2m) based polynomial multiplier and reduction. Here, multiplication follows a hierarchical strategy having parallel processing paths. Figure 11 shows the architecture of proposed Hybrid multiplier for unified NIST standard curves over GF ($2^{163}$). The architecture combines CM and KM on unique optimum point as stated in section III. First stage of architecture is a cascading structure using 2/4/8-bits onwards up to the finalized optimum point i.e. 41 for chosen NIST standard curve as in table 4.

Similarly, first, we look for resource estimations for B-283 and B-571 and then extract the optimal m for them. Further, to generalize the approach for NIST curves reveal stage I for B-283 and B-571, the stagewise sequences are listed, it is observed that apart from B-571, all other curves require a maximum of 2 steps for KM multiplication, excluding the CM case. However, for B-571, 3 steps are required for KM multiplication. This suggests that for higher-degree polynomial multiplications, more than 3 steps may be needed for the KM algorithm listened in resource estimation Table 4.

In the figure 9 the arrows indicate the dataflow with bus width. Final stage is reduction using irreducible polynomial confining the stage II {2m-2} bits output reduced to {m-1} bits, for GF ($2^{163}$) it is 163-bits. Inset shows the reduction using shifting process. The process has two steps, in step 1 P(x) reduces the coefficient with degree > $x^{163}$, and step2 is applied only when the coefficient with ($x^7$, $x^6$, $x^3$) degree exits. The step 2 usually requires zero padding.

## IV. Results and Discussion

This architecture uses a balanced pipeline structure to optimize different multiplication algorithms with a reduction scheme. The architecture for the proposed hybrid approach reduces intermediate registers and gate count significantly improving the routing paths and reducing control logic requirements. The approach leads to desired hardware performance outperforming the existing approaches for the modular multiplier. The architecture in Figure 9 is implemented with digital logic for unified NIST standard curves estimating the resources on the Xilinx AMD zynq-7 series reconfigurable platform as depicted in Table 5. Here the approach utilizes the optimum m for stages I and II as depicted in Table 4 and uses the standard irreducible polynomial for final reduction.

### A. Performance Analysis

This section discusses the implementation results of the proposed method and compares it with other relevant works in this field. For designing a hybrid multiplier shown in Figure 9, using CM and KM, based on our analysis, we observed that the CM exhibits greater efficiency than the KM for small operand sizes, and KM is better for larger bit sizes. In the case of B-163, CM is applied up to 41 bits operand sizes which enhances area efficiency.

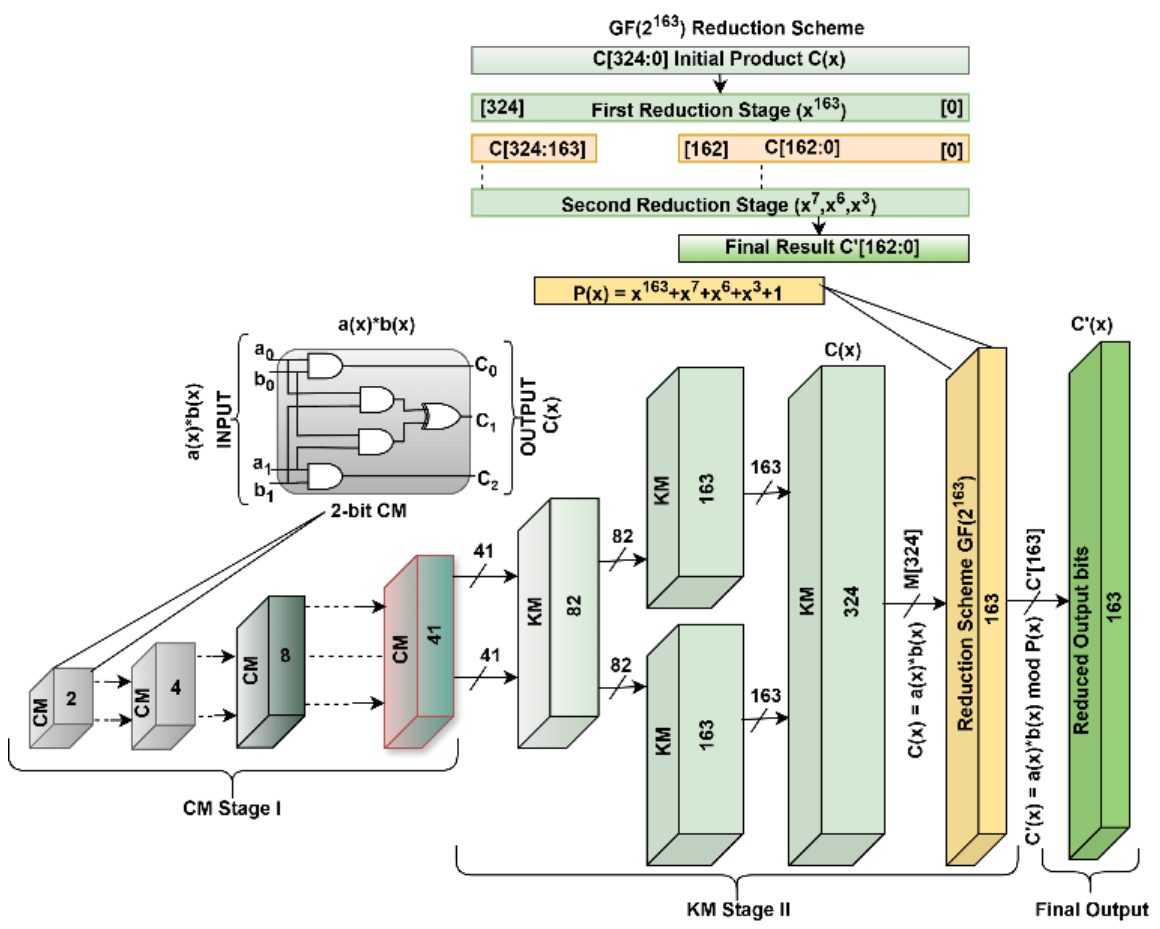

Fig. 9. Architectural diagram to device hybrid multiplication technique combining CM as stage I & KM as stage II, with NIST standard irreducible polynomial over GF (2163) as the final stage.

The CM demonstrates around 0.14% improvement in area and 9.39 % in delay thereby overall 9.55% improvement in ADP in comparison to 41-bit KM. Utilizing CM for Stage I and later KM for Stage II the proposed Hybrid Modular multiplier for constructing a 163-bit multiplier.

Table 5: Resource Estimation of the proposed hybrid design for NIST Standard curve Virtex-7

| Operand Size | LUT | Delay (ns) | ADP |
|---|---|---|---|
| 163 | 6812 | 13.307 | 090860.2 |
| 233 | 10787 | 13.387 | 144405.6 |
| 283 | 14508 | 12.280 | 178158.2 |
| 571 | 44550 | 16.473 | 733872.2 |

The performance comparison between KM and CM implementations reveals distinctive patterns across different operand sizes, demonstrating varying trade-offs in resource utilization, delay, and overall efficiency. However, this efficiency gap narrows considerably as operand size increases, eventually leading to efficiency gains for larger operands, with KM showing a 2.06% improvement at size 82 and a more substantial 13.01% reduction at size 163. These combined observations indicate that KM implementation, despite its consistent delay overhead, becomes increasingly advantageous for larger operand sizes in terms of both resource utilization and overall efficiency. This makes it particularly suitable for applications requiring larger multiplications, while CM remains more efficient for medium-sized operands. The clear transition points in efficiency metrics provide valuable guidance for selecting the appropriate implementation based on specific operand size requirements. A comprehensive comparison between CM, KM, and our proposed hybrid design shown in Table 5, for operand size 163 reveals significant improvements across all performance metrics.

In terms of resource utilization, from the Table 3a &3b, the hybrid design demonstrates remarkable efficiency in LUT usage, requiring 6,812 LUTs compared to 9,982 for CM and 7,762 for KM implementations. This represents a substantial reduction of 31.76% compared to CM and 12.24% compared to KM, indicating superior resource optimization. The delay performance analysis shows equally impressive results. The hybrid design achieves a delay of 13.307 ns, significantly outperforming both CM (18.129 ns) and KM (20.282 ns) implementations. This translates to a 26.60% reduction in delay compared to CM and a 34.39% improvement over KM. Notably, while KM shows an 11.87% higher delay than CM, our hybrid design successfully overcomes this limitation. The Area-Delay Product (ADP) metrics further underscore the hybrid design's efficiency. With an ADP of 90,860.2, the hybrid implementation achieves a remarkable 49.79% improvement over CM (180,963.68) and 42.28% over KM (157,428.88). This substantial reduction in ADP demonstrates that our hybrid approach successfully optimizes both area and delay characteristics simultaneously, rather than trading one for the other. These results validate the effectiveness of our hybrid design strategy, showing that it successfully combines the strengths of both CM and KM approaches while mitigating their respective drawbacks. The Hybrid design shows superior performance across all metrics (LUT, Delay, and ADP) compared to both CM and KM implementations for operand size 163, and overall efficiency demonstrates robustness including the power estimation graph shown in Fig. 10. Also, the well-balanced nature of our hybrid implementation.

Fig. 10. shows a power estimation comparison across different components for what appears to be NIST Standard curve Vertex-7 implementations. The graph compares four different versions/implementations labeled as B-163, B-233, B-283, and B-571. The power consumption is measured across several categories, signals - B-571 shows the highest value at around 2.5-3 units, Logic has a similar pattern to Signals, with B-571 being the highest, I/O All implementations show relatively low values, with slight variations, Dynamic power - B-571 shows significantly higher consumption at around 6 units, Static Power - Very low values across all implementations, Total on-chip power - B-571 shows the highest total power consumption at about 6.5-7 units. The B-571 implementation consistently shows the highest power consumption across most categories, while B-163 generally shows the lowest. The graph suggests that as the bit-length increases (from 163 to 571), the power consumption generally increases as well, particularly noticeable in the dynamic power and total on-chip power measurements. The data appears to be presented in a bar chart format with different colored bars representing each implementation variant, and the y-axis seems to be measuring power units (though the specific unit of measurement is not indicated in the image).

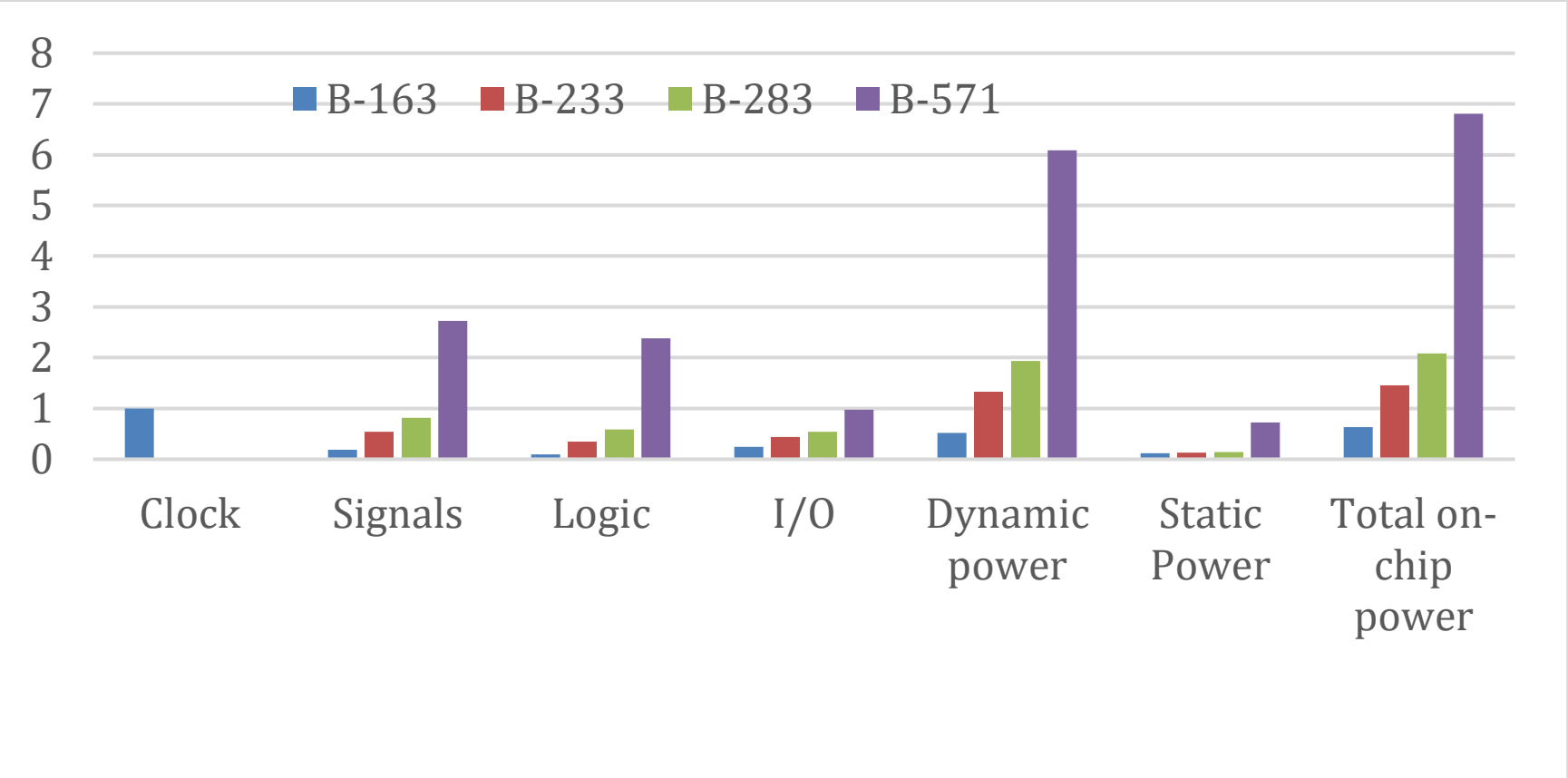

Fig. 10. Power estimation of the proposed design for NIST Standard curve Virtex-7

This diagram shown in Fig. 11. the architecture of a 163-bit hybrid KM implementation. The design consists of four main blocks connected in sequence: 1. Clocking Wizard manages clock signals (clk_in1_0 to clk_out1), 2. VIO (Virtual Input/Output): Handles probe inputs/outputs with 32-bit width, 3. Karatsuba Multiplier: Core computation block showing RTL implementation, 4. ILA (Integrated Logic Analyzer): Monitors and analyses signals, The data flow moves left to right, with clock signals and probe data being processed through the KM, and results being analysed by the ILA block. Each connection shows specific bit-widths and signal names, indicating a detailed hardware implementation for 163-bit multiplication.

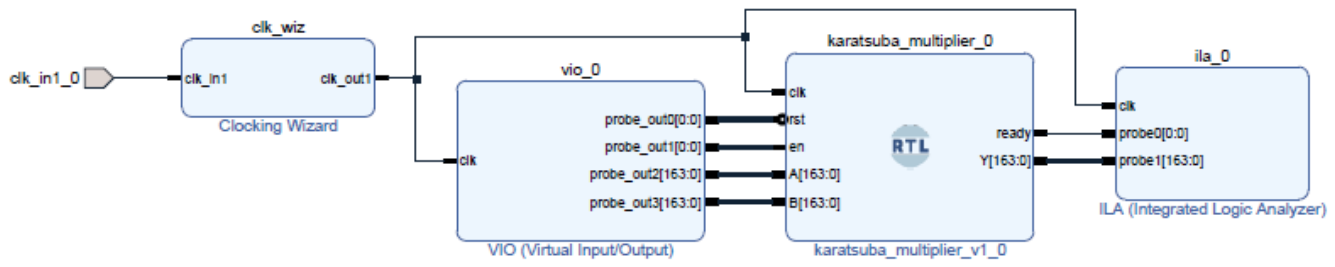

Fig. 11. Verification Architecture of hybrid modular multiplier using ILA and VIO Infrastructure

The verification architecture for hybrid modular multiplier implements a comprehensive testing framework utilizing Integrated Logic Analyzer (ILA) and Virtual Input/output (VIO) infrastructure. The design incorporates a Clocking Wizard for precise timing control, enabling synchronized operation of the verification components. The VIO module facilitates dynamic input vector generation and runtime parameter modifications, allowing interactive testing of the multiplier under various conditions. The ILA core provides real-time signal capture capabilities, enabling detailed monitoring of internal states and critical signal paths.

The architecture employs multiple probe points strategically placed to monitor key signals, including probe out and probe in interfaces. These probe points enable continuous observation of the multiplier's operational states and intermediate results. The ready signal path ensures proper handshaking between the verification infrastructure and the core multiplier logic. This integrated approach allows for thorough validation of the multiplier's functionality, performance characteristics, and timing constraints. The verification system supports both static and dynamic testing methodologies, enabling rapid detection of potential implementation issues and validation of the multiplier's mathematical correctness. The real-time debugging capabilities facilitate immediate feedback during testing phases, significantly reducing the verification cycle time and improving the overall quality assurance process.

### B. Comparison with Existing Research work

The proposed hybrid CM-KM multiplier demonstrates significant performance improvements across various FPGA platforms and operand sizes. Our analysis reveals notable enhancements in three critical aspects: resource utilization, delay performance, and overall efficiency. we can compare our work with existing research work in terms of resource utilization shown in Table 6.

In terms of resource utilization, for operand size m=163, our hybrid design achieves remarkable LUT optimization. It requires only 6,812 LUTs, representing a 39.82% reduction compared to the [37] Bit-Parallel implementation on Artix-7 (11,320 LUTs) and a substantial 71.59% reduction compared to the Montgomery implementation on Virtex-6 (23,977 LUTs) [46]. The design also shows a 12.51% improvement over the LFSR implementation on Virtex-5 [36].

Table 6: Comparing FPGA resource utilization and delay of the proposed multiplier with relevant research work

| Ref. | Device | m | Area | | Delay (ns) | ADP | Reduction | Method | Year |
|---|---|---|---|---|---|---|---|---|---|
| | | | LUT | Slices | | | | | |
| [45] | Spartan 3E | 8 | 62 | 36 | 13.95 | 1367 | No | Modified KM | 2014 |
| [29] | Virtex 4 | 24 | 1018 | 972 | 13.00 | 25870 | No | KM Urdhva | 2015 |
| [36] | Virtex 5 | 163 | 7786 | - | 05.50 | - | Yes | LFSR | 2013 |
| [46] | Virtex 6 | 163 | 23977 | 2061 | 11.70 | - | Yes | Montgomery | 2016 |
| [34] | Stratix II | 163 | - | 1033 | 41.80 | 43179 | Yes | Digit-serial | 2017 |
| [37] | Artix-7 | 163 | 11320 | 3532 | 21.33 | 75337 | Yes | Bit-Parallel | 2018 |
| **Ours** | Virtex 7 | 163 | 6812 | 1335 | 13.31 | 90860 | Yes | Hybrid (CM-KM) | |
| [[47] | Virtex II | 191 | - | 62657 | 45.89 | - | Yes | Montgomery | 2010 |
| [24] | Artix-7 | 233 | 19804 | 1147 | 08.29 | 173683 | Yes | Overlap-free | 2021 |
| [48] | Virtex 4 | 233 | 36812 | 21195 | 07.17 | 3702090 | Yes | Bit-Parallel | 2020 |
| **Ours** | Virtex 7 | 233 | 10787 | 1299 | 13.39 | 144405 | Yes | Hybrid (CM-KM) | |

For larger operand size m=233, the efficiency gains are even more pronounced, with our design using 10,787 LUTs, achieving a 45.53% reduction compared to the Overlap-free design and a 70.70% reduction compared to the Bit-Parallel implementation [24], [48]. And the delay performance analysis reveals competitive results across different operand sizes. At m=163, our design achieves a delay of 13.31ns, showing a 37.60% improvement over the Bit-Parallel Artix-7 implementation. While this is 68.21% slower than the LFSR implementation, it offers better resource utilization as a trade-off. For m=233, the design maintains consistent performance at 13.39ns, demonstrating remarkable stability across different operand sizes, though with some trade-offs against Overlap-free and Bit-Parallel implementations. The Area-Delay Product (ADP) metric demonstrates the overall efficiency of our design. For m=163, the design achieves an ADP of 90,860, showing balanced resource-speed trade-offs compared to other implementations. The efficiency becomes more pronounced at m=233, where our design achieves an ADP of 144,405, representing a 16.86% improvement over the Overlap-free implementation and a remarkable 96.10% improvement compared to the Bit-Parallel implementation. These results demonstrate that our hybrid design successfully balances resource utilization and performance metrics, making it particularly suitable for applications requiring efficient FPGA resource usage while maintaining competitive delay characteristics. The proposed hybrid CM-KM multiplier demonstrates several distinctive advantages that establish its effectiveness for cryptographic applications. A primary strength lies in its scalability, where the design maintains consistent performance characteristics across different operand sizes. The transition from m=163 to m=233 shows proportional resource utilization while maintaining stable delay metrics, indicating robust scaling capabilities.

The design achieves remarkable efficiency through its balanced approach to performance optimization. While significantly reducing resource requirements, it avoids severe compromises in delay performance, demonstrating that resource optimization and speed can be simultaneously addressed. This balance is maintained across various FPGA platforms, highlighting the design's versatility and practical applicability. Implementation-wise, the hybrid design successfully incorporates reduction techniques while maintaining competitive performance metrics. The effective integration of CM and KM methodologies results in optimized resource utilization without significantly impacting operational speed. This balanced approach makes the design particularly suitable for applications where efficient resource usage is crucial, but performance cannot be significantly compromised.

These characteristics make our hybrid design an excellent choice for practical cryptographic implementations, offering a well-balanced solution that effectively addresses the traditional trade-offs between resource utilization and performance. The design's ability to maintain these advantages across different FPGA platforms and operand sizes demonstrates its robustness and practical viability in real world applications. The comprehensive analysis demonstrates that our proposed hybrid architecture offers a viable alternative to existing implementations, particularly in scenarios where balanced performance and resource utilization are primary concerns.

## V. Conclusion

In this paper, we introduce a novel approach that extends the hybrid multiplication (CM-KM) technique to modular multipliers. One of the most appealing aspects of the new algorithm is that it allows designers to arbitrarily choose the degree of the defining irreducible polynomial. Additionally, this new field multiplier results in architectures with significantly reduced gate complexity compared to conventional methods. Moreover, the new multiplier facilitates highly modular architectures, making it particularly suitable for VLSI implementations. Our proposed hybrid (CM-KM) modular multiplier design demonstrates notable improvements in several key metrics when compared with existing implementations across different FPGA platforms and operand sizes. The analysis can be categorized into three main aspects: 1. Resource Utilization Analysis: For m=163 implementations, our hybrid design achieves significant resource optimization: Requires only 6,812 LUTs compared to 11,320 LUTs in Bit-Parallel (Artix-7) implementation, representing a 39.82% reduction, shows 71.59% LUT reduction compared to Montgomery implementation on Virtex-6 (23,977 LUTs), Utilizes 12.51% fewer LUTs than LFSR implementation on Virtex-5 (7,786 LUTs). For m=233 implementations demonstrate substantial improvement with 10,787 LUTs compared to 19,804 LUTs in Overlap-free design (45.53% reduction), 36,812 LUTs in Bit-Parallel implementation (70.70% reduction). 2. Delay Performance: The delay characteristics show competitive performance, For m=163: Our design achieves 13.31ns delay, showing, 37.60% improvement over Bit-Parallel Artix-7 implementation (21.33ns), Competitive performance compared to Montgomery implementation (11.70ns), 68.21% slower than LFSR implementation (5.50ns), but with better resource utilization, For m=233: Maintains consistent performance with 13.39ns delay, Trade-off against Overlap-free (8.29ns) and Bit-Parallel (7.17ns) implementations, Demonstrates remarkable consistency across different operand sizes. 3. Area-Delay Product (ADP) Analysis: The ADP metric reveals the overall efficiency of our design: For m=163, Achieves 90,860 ADP compared to 75,337 in Bit-Parallel implementation, shows better efficiency than Digit-serial implementation (43,179), Demonstrates balanced resource-speed trade-off. For m=233, Achieves significant improvement with 144,405 ADP, 16.86% better than Overlap-free implementation (173,683), 96.10% improvement over Bit-Parallel implementation (3,702,090).

### ACKNOWLEDGMENT

This research was supported through CSIR-HRDG, with additional support and the resources of CSIR-CEERI and the AcSIR academic program.